# Digital Audio Processing Tools for Music Corpus Studies
Submitted to: *Oxford Handbook of Music and Corpus Studies*
Johanna Devaney

**Abstract**

Digital audio processing tools offer music researchers the opportunity to examine both non-notated music and music as performance. This chapter summarises the types of information that can be extracted from audio as well as currently available audio tools for music corpus studies. The survey of extraction methods includes both a primer on signal processing and background theory on audio feature extraction. The survey of audio tools focuses on widely used tools, including both those with a graphical user interface, namely Audacity and Sonic Visualiser, and code-based tools written in the C/C++, Java, MATLAB, and Python computer programming languages.

## 1 Introduction

Early work on music corpus studies focused on symbolic music data such as a score, and these data continue to dominate to the present day. For a detailed discussion of these, see the chapter by Eamon Bell in this volume for a summary of symbolic toolkits and the double special issue on corpus studies in Music Perception for examples of work on symbolic data (Temperley and VanHandel 2013, VanHandel and Temperley 2014). There has been a relatively small, but growing, body of research on audio corpus studies, influenced by developments in audio processing tools. This chapter summarises both the types of musically-relevant information that can be extracted from musical audio signals and some of the most commonly used tools for extracting this information.

The general approach to audio corpus studies is similar to symbolic studies, insofar as the audio data can be converted into some form of symbolic representation, which may be a transcription or set of features, which may be linked to a score or not. Both symbolic and audio data can be characterised by discrete events (typically notes), when the appropriate transcription algorithms are available for audio processing, or they can be treated as signals (if you represent the score data as a piano roll) or as sets of features, although the latter is more common in digital audio processing. Treating the music data as a signal allows for summarisation at larger time scales than the note-level, which can be useful for smoothing information and looking at longer-scale structures. What is consistent across these approaches is the use of a two-step process where features are extracted from the data and then summarised with a statistical technique (be it descriptive or inferential).

Extracting features from musical audio signals is more challenging than from symbolic signals (although there are indeed challenges in creating clean and accurate symbolic representations of printed scores). Corpus studies of musical audio do, however, facilitate different research questions than those that can be posed of symbolic corpora. This is because, in addition to allowing the pitch content of non-notated music to be studied in a way similar to notated music, audio corpus studies allow for a range of musical attributes to be examined that are not available in, particularly score-based, symbolic representations, such as micro-timing, pitch-inflections, and dynamics and timbral variations.

Much of the current work on audio signals has required a significant amount of intervention on the part of the researcher, sometimes to the degree of manually transcribing or annotating the entire corpus. This was common both in early ethnomusicology work on recordings (see the chapter by Peter Savage in this volume and the recent article by Panteli, Benetos, and Dixon (2018) for a history of some of these corpora), as well as more recent work on film (Richards 2016), performance practice in Western art music (Cook 2007, Leech-Wilkinson 2010, Timmers 2007), and popular musics (Biamonte 2014, Easley 2015, Richards 2017, De Clercq 2017, De Clercq and Temperley 2011). Some of this recent work has taken advantage of tools developed in the music information retrieval community that allow for automatic or semi-automatic analysis of the musical signal to study both jazz (Frieler et al. 2016, Abeber et al. 2017), and a range of world musics (Serra 2014, Holzapfel, Krebs, and Srinivasamurthy 2014), including Hindustani (Srinivasamurthy et al. 2017) and Malian (London, Polak, and Jacoby 2017). Thus, there currently appears to be a divide between the types of information that can be reliably automatically extracted from audio signals and the types of questions that interest more traditional music researchers. In light of this, this chapter considers both tools that facilitate manual and semi-automatic annotations through a graphical user interface and code-based tools for automatic analysis of audio signals. Before surveying these tools, however, we begin with some background information on audio signal processing.

## 2 Features

Features are numerical descriptors extracted from the audio signal. They can be calculated at different time scales and, when extracted from time-frequency representations, across different frequency ranges. Musical audio-processing tools typically work by extracting or estimating information about the audio signal in millisecond-scale chunks, typically referred to as frames. These frame-wise descriptors are then analysed in order to describe longer-scale structures in the audio. Musically-meaningful longer-scale structures may be at the note-, beat-, bar-, phrase-, section-, or song-levels and can summarise timing, pitch, loudness, and timbral characteristics of the audio. This section provides a brief introduction to the fundamental concepts underlying audio analysis, for a more in-depth discussion, please see Gold, Morgan, and Ellis (2011) and Müller (2015).

### 2.1 Signal Processing basics

Most of the signal processing algorithms used in the tools discussed in this chapter begin with a time-frequency analysis of the audio signal. In order to understand time-frequency analysis, it is useful first to consider the physical nature of sound. Sound waves move through the air as changes in sound pressure, which can be represented in the time-domain, which shows the signal's amplitude as a function of time. An example of a time-domain representation of a vocal signal is shown in Figure 1 (for the score excerpt reproduced in Figure 2).

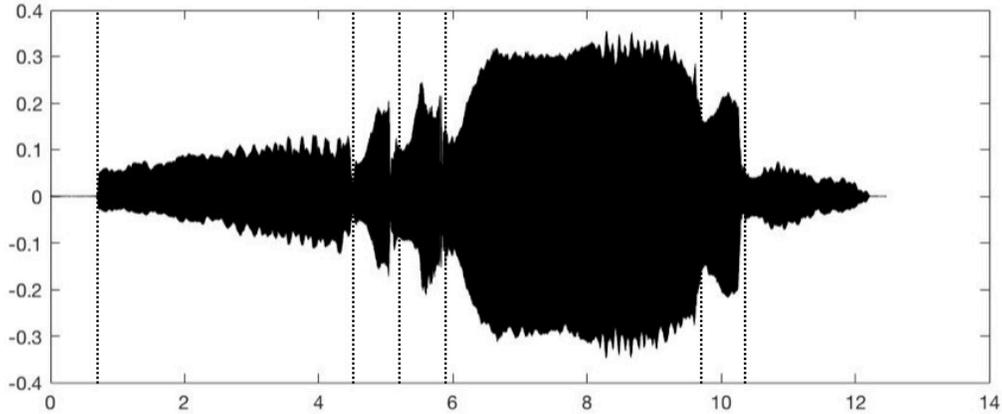

Figure 1: Time domain representation of an audio recording of the opening phrase of Schubert's "Ave Maria". The vertical dotted lines indicate the boundaries of the notes in the score in Figure 2.

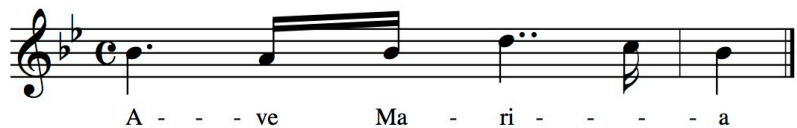

Figure 2: Score of the opening phrase of Schubert's "Ave Maria", a recording of which is represented in the time-domain in Figure 1 and in the time-frequency domain in Figure 3.

Vocal signals, along with most non-percussive instruments, produce waveforms that are periodic in that the sound pressure varies with a repeated pattern. Each repetition of this pattern is called a cycle and the number of cycles per second is referred to as frequency. The frequency content of each sound is not just the pitch that you perceive, but also a series of higher frequencies, generally in whole-number ratios to the fundamental/perceived pitch but also some inharmonicities, that creates the sound's timbre. Thus, we can distinguish between two different instruments, say a clarinet and a violin, playing the notes with the same onset and duration at the same pitch with the same loudness.

In the early 19th century Joseph Fourier proved that almost any periodic signal can be represented by a sum of sinusoids. Fourier-based time-frequency analysis provides frequency information about both the harmonic and non-harmonic sinusoids that a sound can be decomposed into and the amplitude of these sinusoids. Fourier's theorem has been implemented as the Discrete Fourier Transform (DFT), which allows for the analysis of discrete, rather than continuous signals. The frequency resolution of the DFT is related to the length of the signal being analysed. There is a trade-off between the temporal acuity of the DFT and the lowest analysable frequency since lower frequencies require longer observations, which blur temporal events, which is important to consider when selecting a window size. Since the DFT is typically modelling the signal with multiples of a sinusoid lower than the fundamental frequencies of the notes in the signal, the DFT is able to capture both harmonic and non-harmonic frequency components of the musical sounds in the signal.

A series of DFT analyses can be stacked into a spectrogram. This is done by applying the DFT to chunks (or windows) of the signal with a small temporal offset between the start of each chunk, referred to as the hop size (or window increment). The size of the chunk is referred to as the width or window size. A spectrogram provides a 2D matrix containing the amplitudes of the frequency components present in the signal during each window, with one axis divided into frequency bins and the other into time segments. In a visual representation of a spectrogram, the amplitude of the frequency components at each point in time are represented by colour. See Figure 3 for an example of a spectrogram.

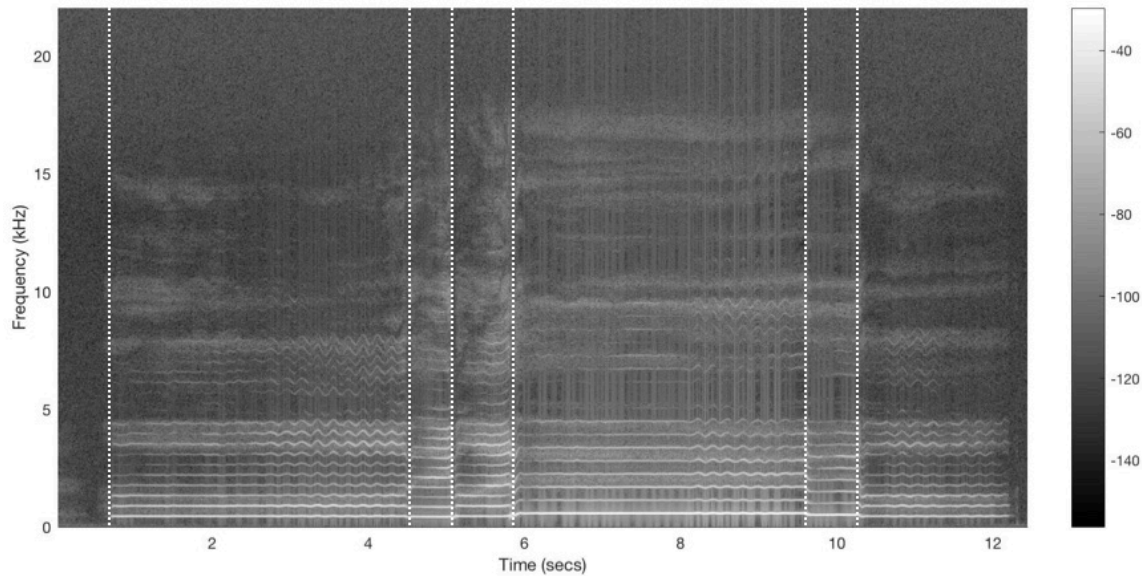

Figure 3: Spectrographic (time-frequency) domain representation of an audio recording of the opening phrase of Schubert's "Ave Maria". The vertical dotted lines indicate the boundaries of the notes in the score in Figure 2.

The spectrogram described above is one of several time-frequency representations that one can produce from audio. The size of the frequency bins in a DFT spectrogram are all the same. In a *constant-q spectrogram*, the frequency bin size increases in proportion to frequency so that the lower frequencies, which human hearing has better frequency-resolution for, can be represented with greater acuity. In a *mel spectrogram*, the sizes of the frequency bands are more closely linked to the frequency mapping of human hearing in that the frequency mapping is linear below 1 kHz and logarithmic above. The *bark spectrogram* is the same, but uses the slightly different bark frequency scale, which makes its linear-to-logarithmic transition around 500 Hz. It is useful to understand the differences between these representations, as some tools can apply their audio processing algorithms to several representations, which can influence how well the algorithm performs on different combinations of instruments or different styles of music.

For monophonic music, the relationship between the time- and, particularly, the time-frequency-domain representations of the audio can be easily understood. For polyphonic music, however, where more than one instrument or voice is present or when an instrument plays multiple notes at the same time, it is more complicated, both for humans and for computers. Full polyphonic-transcription is still an unsolved research task, so in its absence, currently available audio processing tools extract features about the musical audio rather than transcriptions of the notes. These features capture information not only related to timing and pitch but also timbre and dynamics, and thus they are potentially more useful than a transcription. These features are discussed in detail in Section 2.2.

Filters are often used as a pre-processing step in audio analysis as they reduce the amount of information in the signal. For example, if one is interested in analysing melodies, the use of a low-pass filter would be beneficial as it would remove the high-frequency content that is not relevant to the analysis task. The four basic types of filters are high-pass, where frequencies above the specified threshold pass through, low-pass, where frequencies below the threshold pass through, band-pass, where frequencies between two specified thresholds pass through, and band-stop, where frequencies below the lower specified threshold and above the higher specified threshold pass through. The roll-off is an important parameter in filters as it defines the sharpness of the edges of the filter.

## 2.2 Types of Features

Features are often divided into lower- and higher-level features. Lower-level features describe the acoustic properties of the signal while higher-level features describe musical events like notes, chords, or instrumentation. Sometimes (e.g., Bello and Pickens(2005)) the term 'mid-level' has been used to describe features that contain more musically-relevant information than low-level features like power or spectral characteristics, but do not explicitly describe musical events in the signal. Examples of this are fundamental frequency traces, chroma vectors, and onset detection functions. Since this term is not consistently used, however, this chapter will categorise features as either low- or high-level, as summarised in Table 1 and described in Sections 2.2.1 and 2.2.2.

| Feature type | Low-level features | High-level features |
|---|---|---|
| Loudness | RMS<br>Intensity | - |
| Pitch | Fundamental frequency (F0)<br>Chroma vector | Melody<br>Chords<br>Key<br>Tuning |
| Timbre | MFCCs<br>Synchronicity<br>Spectral descriptors<br>Zero-crossing rate | - |
| Timing | Onset detection function | Onsets<br>Offsets<br>Beats<br>Downbeats<br>Tempo |
| Mixed | - | Note tracking<br>Segments |

Table 1: Table of the most commonly implemented features in the audio processing tools surveyed in Section 3.

### 2.2.1 Low-level features

In monophonic audio, frame-wise *fundamental frequency* (the lowest frequency partial in a periodic waveform, sometimes used synonymous with pitch and referred to as F0) can be estimated. F0 estimation analyses the periodicity in the signal or the frequency content in order to identify the F0, which is typically synonymous with the sounding pitch, for each frame of audio. Some tools provide continuous pitch estimates while others provide summaries of each note. In polyphonic music, *chroma* features can be used as a weak proxy for pitch. Chroma vectors summarise how much energy there is in each of the 12 pitch classes used in the Western scale and can be used to estimate higher-level descriptors of pitch, such as chords and musical key. Both F0 and chroma estimates can be visualised in 2-D representations similar to spectrograms, typically called the *F0gram* and *chromagram*, respectively. There are also a range of loudness-related descriptors that can be extracted from audio signals. Like pitch, loudness is a perceptual construct, one that is based on both amplitude and frequency information within the signal. While some audio processing toolkits provide estimates of perceptual loudness, others provide low-level descriptors that quantify the amount of physical energy in the signal. The most common way to describe the latter is with *root mean square (RMS) pressure*, the square root of the mean power values for each frame in the signal or with *intensity*, the product of sound pressure level and air particle velocity.

While theory about timbre is less comprehensive than pitch or loudness, some audio processing tools offer the ability to calculate timbre descriptors that have been found to be meaningful by psychologists. Some descriptors, such as *spectral centroid*, emerged from the results of psychological tests seeking acoustic correlates of multidimensional scaling embeddings. These embeddings approximate the results of listening tests where participants were asked to listen to notes from different musical instruments and to rate the similarity between them (Grey, 1977). From related literature, there are a range of descriptors that can be divided into two groups, those that characterise the amount of *synchronicity* within the spectrum (such as *inharmonicity* and *tonality*) and those that describe trends in the *distribution of energy in the signal (such as spectral centroid, spectral flatness, spectral flux, spectral roll-off, spectral spread*, *kurtosis*, etc., for details on these descriptors see Peeters et al. (2011)). Other descriptors, such as *mel frequency cepstral coefficients* (MFCCs), were designed based on what is known about pitch processing in the peripheral auditory system (Logan, 2000). In contrast, some descriptors are purely engineered, such as *zero crossing rate*, which is the rate at which the signal moves from positive to negative amplitude in the time domain and is useful for identifying percussive sounds.

The temporal energy characteristics of the signal across the spectrum can provide important clues for discerning event timing information. The standard approach to event onset detection, detailed below, makes use of low-level *onset detection functions*, which capture novel events in the audio signal. Onset detection functions can be calculated from a range of parameters, including high-frequency content, spectral difference, or phase deviation (Bello et al. 2005).

### 2.2.2 High-level features

*Note segmentation* is typically an important step in dividing up the musical signal for further analysis and requires estimation of the timing of the start (*onset* detection) and stop (*offset* detection) of each note. An onset is an estimate of the point at the start of the note that best characterisesthe transient, which Bello et al. define as "short intervals during which the signal evolves quickly in some nontrivial or relatively unpredictable way" (Bello et al. 2005) [p. 1036]. The three stages of typical onset detection approaches are preprocessing, reduction, and peak picking. Offsets can be estimated from decays in amplitude, although less work has been done on this task and the results may be less reliable. Onset and offset detection can be done as a preliminary step before doing further analysis of the signal or, for monophonic signals, it can in informed by pitch analysis. One of the most significant challenges in onset detection is accounting for the varying characteristics of the onsets of different types of instruments, while percussive onsets (such as the ones produced by drums or the piano) can be easily characterisedby amplitude information, non-percussive onsets (particularly those for unfretted string instruments or vocals) are not well characterised by it. Thus, multiple algorithms have been developed and are implemented in available audio processing tools to address the breadth of the problem. A separate but related task is *beat detection*, which attempts to find the metrically significant onsets in a signal and is important for the task of *tempo estimation*.

Pitch tracking and note tracking are related tasks, although note tracking includes note segmentation. In monophonic music, pitch tracking is synonymous with an F0 trace but in polyphonic music, pitch tracks must be estimated from other features. Another pitch-related task is *melody extraction*, sometimes referred to as *predominant melody extraction*, where, typically, a *pitch salience* function is designed that can assist in selecting *pitch contours* from a frequency-domain representation of the audio signal. Pitch contours are trajectories that encompass one or more notes. Further processing, either heuristic or statistical (or both), is then applied to the pitch contours in order to identify those that belong to the melody. Chords are typically estimated from chroma features using a combination of heuristics and statistical learning. A combination of feature types (loudness, pitch, timbre, and timing) can be used to make estimates of the boundaries between musical segments. A secondary step that can be taken is to calculate the similarity between segments and automatically generate labels.

**3 Audio Processing Tools**

**3.1 Tools with Graphical User Interfaces**

Audacity (Mazzoni and Dannenberg 2000) is primarily an audio-editing tool, but it also has annotation capabilities. Manual annotation can be done by listening to the audio file in combination with either a waveform or spectrogram representation of the audio, examples of these are shown in the upper and lower plots of Figure 4, respectively. Annotations are made in a 'label track' and can specify the starting and ending points of musical events, such as notes, labelled with text. The example in Figure 4 shows the six notes of the audio temporally delineated and labelled with the sung text. The 'label track' can be exported into a tab-separated text file, which can be imported in software like Excel or read in a range of programming languages. Audacity also supports analytic plug-ins, which can be used for automatic annotation. The built-in plug-ins primarily focus on audio editing, but the beat and silence finders may be useful for audio corpus studies. Audacity also supports Vamp plug-ins, which are discussed in detail below.

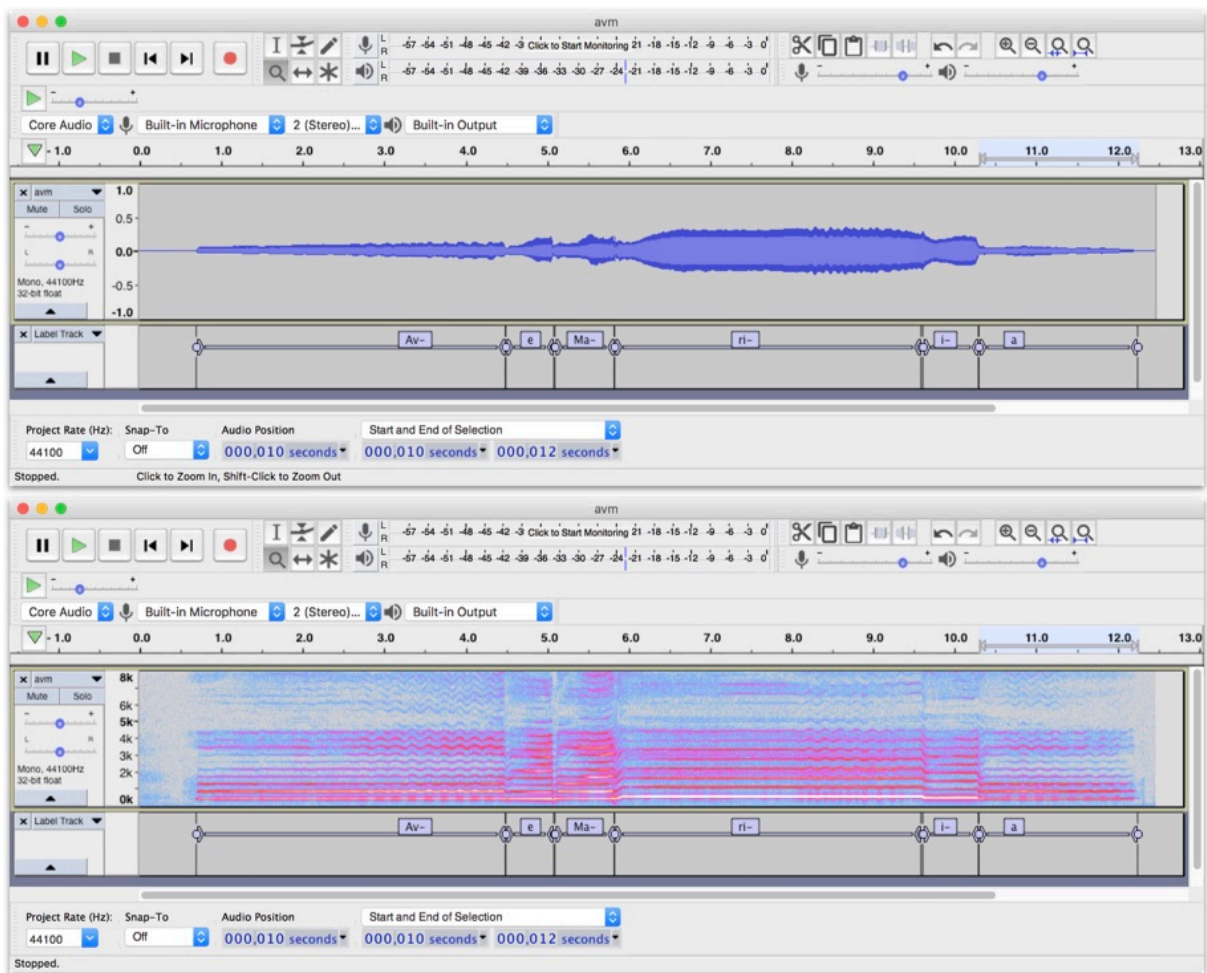

Figure 4: Visualisations of an audio recording in Audacity of the opening phrase of Schubert's "Ave Maria". The top panel is a time-domain waveform and the bottom is a spectrogram. The label track in both panels correspond to the notes in the score in Figure 2

A more developed annotation tool is Sonic Visualiser, which was developed and is maintained by researchers in the Centre for Digital Music (C4DM) at Queen Mary, University of London (Cannam et al. 2006). Since its release in 2006, it has become the mainstay of audio research by musicologists and theorists (Abdallah et al. 2017), due to both its graphical user interface and the wide range of analysis tools available via plug-ins. Sonic Visualiser reads Vamp plug-ins (Cannam 2009), which were also developed at C4DM. Vamp plug-ins can also be used, as noted above, in Audacity as well as the command-line tool Sonic Annotator (Cannam et al. 2010), which is discussed below. This section provides an overview of the capabilities of Sonic Visualiser and VAMP plug-ins, for a detailed tutorial on how to use Sonic Visualiser for musical corpus studies see Cook and Leech-Wilkinson (2009).

The visualisation component of Sonic Visualiser allows for audio files to be displayed in a variety of ways. In addition to the time-domain waveform (shown in the top-left panel of Figure 4) and spectrogram (top-right panel), discussed above, Sonic Visualiser can also visualise audio as a melodic range spectrogram (bottom-left panel), which focuses on the likely range of F0 frequencies, or as a peak frequency spectrogram (bottom-right panel), which focuses on the loudest frequencies (which may or may not be the melody). The visual representations can be layered on top of one another, along with annotation layers. Annotations may be manually labelled, referencing time instants, notes, regions, text, or images, or they may be automatically generated using VAMP plug-ins.

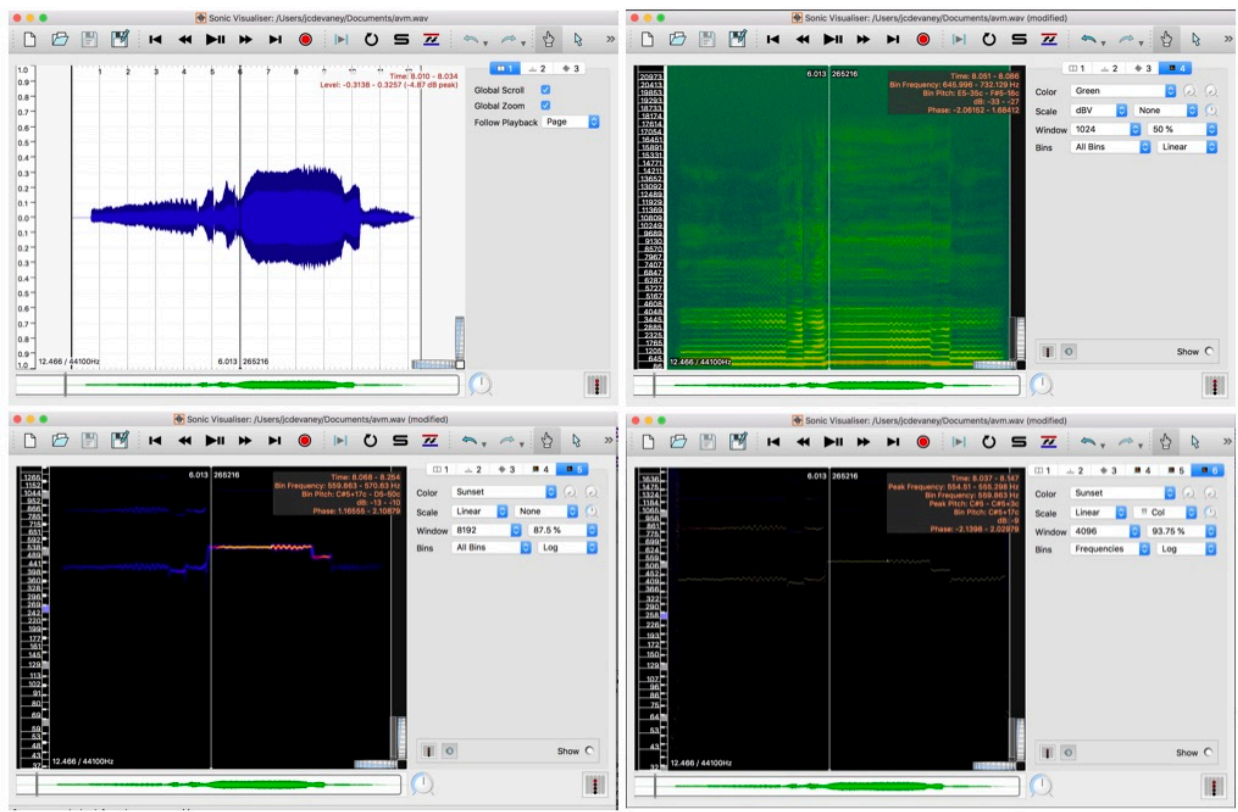

Figure 5: Visualisations of an audio recording in Sonic Visualiser of the opening phrase of Schubert's "Ave Maria", corresponding to the notes in the score in Figure 2. The top-left panel is a time-domain waveform, the top-right panel is a spectrogram, the bottom-left panel is a melody-range spectrogram, and the bottom-right spectrogram is a peak-frequency spectrogram.

VAMP plug-ins allow researchers to release their algorithms in a format usable by people who don't want to deal with code. As noted above, VAMP plug-ins can be used in Sonic Visualiser, Sonic Annotator, and Audacity. C4DM maintains an SDK for researchers to build VAMP plug-ins and they can be written in C++ using this or in Python using the Vampy wrapper. There is also a wrapper, jVAMP, that allows VAMP plug-ins to be run by Java applications (Cannam 2009).

VAMP plug-ins typically provide representations and low-level features that can be used as the basis for further processing and analysis. Some plug-ins group together a range of low- and high-level feature extraction algorithms (e.g., Aubio, BBC Vamp Plug-ins, libxtract Vamp plug-ins, MARSYAS Vamp plug-ins, Mazurka Plug-ins, MIR.EDU and the Queen Mary plug-in set), while others focus on specific tasks, such as beat tracking (BeatRoot INESC Porto Beat Tracker), chord estimation (Chordino), melody extraction (MELODIA), onset detection (OnsetsDS), pitch and note tracking (pYIN), and section segmentation (Segmentino). Table 2 summarises which VAMP plug-ins implement the low- and high-level features discussed in this chapter.

|  | Alicante | AMPACT | Aubio | BBC Plugins | BeatRoot | Cepstral Pitch | Chordino | Chroma Toolbox | Essentia | INESC Porto | jAudio | LibROSA | LibXtract | Madmom | MARSYAS | Mazurka | MELODIA | MIR.EDU | MSFA | MIR Toolbox | OnsetDS | QM Plugins | pYIN | Segmentino | Silvet | Timbre Toolbox | Tony |
|---|---|---|---|---|---|---|---|---|---|---|---|---|---|---|---|---|---|---|---|---|---|---|---|---|---|---|---|
| **Low-level Features** | | | | | | | | | | | | | | | | | | | | | | | | | | | |
| RMS |  | X |  | X |  |  |  |  | X |  | X |  | X |  |  |  |  |  |  | X |  | X |  |  |  |  |  |
| Intensity |  | X |  | X |  |  |  |  | X |  |  |  |  |  | X |  |  |  |  |  |  |  |  |  |  |  |  |
| F0 |  | X | X |  |  | X |  |  | X |  |  | X |  | X |  |  |  |  |  | X |  |  | X |  |  |  | X |
| Chroma |  |  |  |  |  |  | X | X | X |  |  | X |  |  | X |  |  |  |  | X |  | X |  |  |  |  |  |
| MFCCs |  | X | X |  |  |  |  |  | X |  | X | X | X |  | X |  |  |  | X |  | X | X |  |  |  |  |  |
| Synchronicity |  |  |  |  |  |  |  |  | X |  |  |  | X |  |  |  |  |  |  |  | X |  |  |  |  | X |  |
| Spectral Desc |  | X | X | X |  |  |  |  | X |  | X | X | X |  | X | X |  |  | X |  | X |  |  |  |  | X |  |
| Zero-crossing |  |  |  |  |  |  |  |  | X |  | X | X | X |  | X |  |  |  | X |  | X |  |  |  |  | X |  |
| Onset Detect Func | X | X | X |  |  |  |  |  | X |  |  | X |  |  |  |  |  |  |  |  | X |  |  |  |  |  |  |
| **High-Level Features** | | | | | | | | | | | | | | | | | | | | | | | | | | | |
| Melody |  |  |  |  |  |  |  |  | X |  |  |  |  |  |  |  | X |  |  |  |  |  |  |  |  |  |  |
| Chords |  |  |  |  |  |  | X |  | X |  |  | X |  |  |  |  |  |  |  |  |  | X |  |  |  |  |  |
| Key |  |  |  |  |  |  |  |  | X |  |  | X |  |  |  |  |  |  |  | X |  | X |  |  |  |  |  |
| Tuning |  |  |  |  |  | X | X | X |  |  |  |  |  |  |  |  |  |  |  |  |  |  |  |  |  |  |  |
| Onsets | X | X | X | X |  |  |  |  | X |  | X |  | X |  | X |  |  | X |  | X | X | X |  |  |  |  | X |
| Offsets |  | X |  |  |  |  |  |  |  |  |  |  |  |  |  |  |  |  |  |  |  |  |  |  |  |  | X |
| Beats |  |  | X |  | X |  |  |  | X | X |  | X | X | X |  |  |  |  |  | X |  | X |  |  |  |  |  |
| Downbeats |  |  |  |  |  |  |  |  |  |  |  |  |  | X |  |  |  |  |  |  |  | X |  |  |  |  |  |
| Tempo |  |  | X |  |  |  |  |  | X |  |  | X |  | X | X |  |  |  |  | X |  | X |  |  |  |  |  |
| Note tracking | X | X | X |  |  |  |  |  | X |  |  | X |  |  |  |  |  |  |  |  |  | X |  | X |  | X |  |
| Segments |  |  |  |  |  |  |  |  | X |  |  |  |  |  |  |  |  |  | X | X |  | X |  | X |  |  |  |
| **Machine Learning** | | | | | | | | | | | | | | | | | | | | | | | | | | | |
| PCA |  |  |  |  |  |  |  |  |  | X |  |  |  |  | X |  |  |  |  |  |  |  |  |  |  |  |  |
| SOM |  |  |  |  |  |  |  |  |  |  |  |  |  |  | X |  |  |  |  |  |  |  |  |  |  |  |  |
| KNN |  |  |  |  |  |  |  |  |  | X |  |  |  |  | X |  |  |  |  | X |  |  |  |  |  |  |  |
| GMM |  |  |  |  |  |  |  |  |  |  |  | X | X |  |  |  |  |  |  | X |  |  |  |  |  |  |  |
| SVM |  |  |  |  |  |  |  |  |  | X |  |  |  |  | X |  |  |  |  |  |  |  |  |  |  |  |  |
| NN |  |  |  |  |  |  |  |  |  | X |  |  | X |  |  |  |  |  |  |  |  |  |  |  |  |  |  |
| HMM |  |  |  |  |  |  |  |  |  |  |  | X |  |  |  |  |  |  |  |  |  |  |  |  |  |  |  |
| CRF |  |  |  |  |  |  |  |  |  | X |  |  |  |  |  |  |  |  |  |  |  |  |  |  |  |  |  |

Table 2: Summary of the low-level features, high-level features, and machine learning algorithms implemented in the audio processing tools discussed in this chapter.

Onset detection VAMP plug-ins include Aubio (Brossier 2006), OnsetDS (Stowell and Plumbley 2007), the Queen Mary plug-in set's onset detector (Duxbury et al. 2003, Stowell and Plumbley 2007, Barry et al. 2005), and the University of Alicante Vamp Plug-ins' onset detector (Pertusa and Iñesta 2009). Some plug-ins, e.g., Queen Mary's plug-in set, provide both onset detection and beat tracking, while others focus exclusively on beat tracking as a separate task: BeatRoot (Dixon 2001) and the INESC Porto Beat Tracker (Oliveira et al. 2012). In contrast, the MIR.EDU plug-in (Salamon and Gómez 2014) calculates both the starting and stopping times of the attack portion of each note (Peeters 2004).

Both pitch and note tracking are available in the Cepstral Pitch Tracker (Cannam et al. 2013) and pYIN (Mauch and Dixon 2014) plug-ins. Higher-level pitch features can also be estimated with VAMP plug-ins, including melody with MELODIA (Salamon and Gómez 2012), and chord estimation with Chordino (Mauch and Dixon 2010). Some plug-ins, such as Silvet Note Transcription (Benetos and Dixon 2012, 2013) and University of Alicante (Pertusa and Iñesta 2012), perform polyphonic transcription, but the underlying algorithms, with accuracy rates of less than 70%, are not robust enough to provide transcriptions sufficiently accurate for most corpus study purposes. The segmentation algorithms, such as the ones in Segmentino (Mauch, Nol, and Dixon 2009, Cannam et al. 2013) and the Queen Mary plug-in set (Levy and Sandler 2008), can be similarly inexact but can serve an exploratory purpose.

Tony is a related tool, built on the code-base of Sonic Visualiser that is designed for annotation of melodies (Mauch et al. 2015), something that is possible to do in Sonic Visualiser but easier and more straightforward in Tony, with everything in a single pane. Also, note segmentation and pitch estimation happen automatically and options can be set through menus. These options include using unbiased timing, which changes the way the F0 analysis is performed, penalising soft pitches, since low-amplitude pitches are likely spurious, applying a high sensitivity for onsets, and dropping short notes. Errors in the automatic estimation can be corrected manually, with options to create, split, delete, and merge notes. Tony can display the waveform, spectrogram, F0 trace, and estimated notes. It also gives the option for sonifying for the F0 trace and estimated notes, in combination with or separate from the original waveform and/or each other.

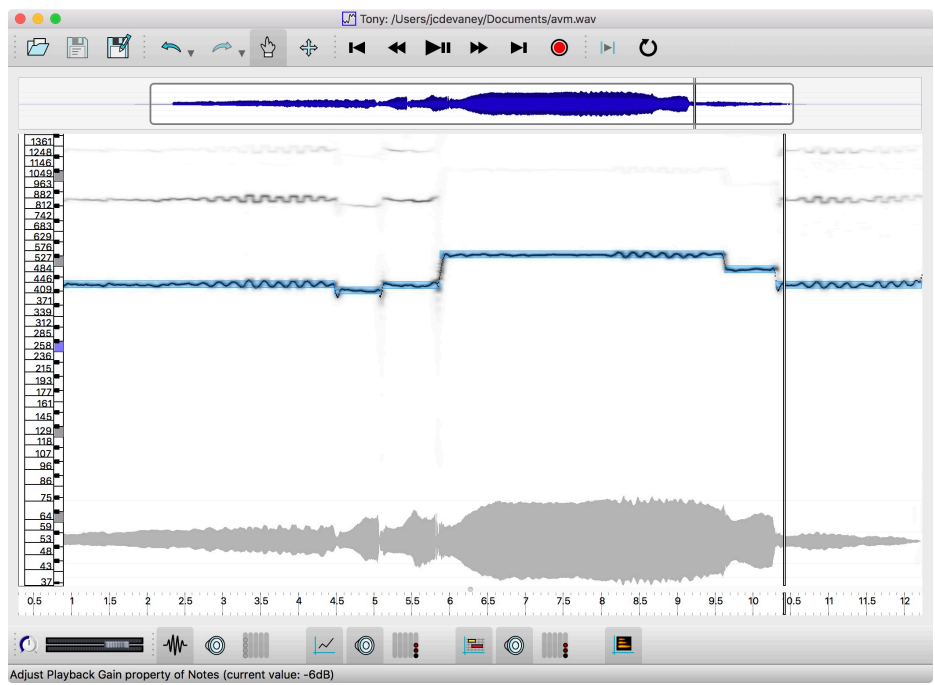

Figure 6: Visualisations of an audio recording of the opening phrase of Schubert's "Ave Maria", corresponding to the notes in the score in Figure 2, in Tony.

Sonic Annotator is another tool developed by C4DM that allows for VAMP plug-ins to be run on the command-line. Whereas Sonic Visualiser is intended for close analysis of an audio file through an interaction of analysis (either manual or automatic) and visualisation using different representations of the audio signal, Sonic Annotator allows a batch of audio files to be processed automatically with VAMP plug-ins. Sonic Visualiser is an excellent tool for learning about and understanding your data, fine-tuning your analytical approach, and generating manual or computer-assisted annotations. If your analytical approach can be accurately executed with VAMP plug-ins (requiring no manual intervention), then Sonic Annotator can be used to process your data more quickly.

## 3.2 Code-Based Tools

There are a large number of code-based tools for audio processing, which also allow for batch processing of audio files using existing algorithms (like Sonic Annotator does). Some of the available code-based tools have been briefly mentioned in the above discussion of VAMP plug-ins. This section covers the most widely used tools in research over the past twenty years. In terms of the specific types of tools: Frameworks, libraries, and toolkits all extract some of the features described above, while frameworks also provide statistical learning tools. One of the benefits of the code-based tools is that tools can be combined. This can be done easily within the same language but even tools written in different languages can be integrated (e.g., by calling MATLAB functions in Python and vice-versa). As with VAMP plug-ins, the specific functionalities of the code-based tools are detailed in Table 2. In addition, Table 3 provides a summary of the tools covered in this section, what language they are written in, whether it is possible to run the code in real-time (which is not possible with any of the GUI-based tools discussed in Section 3.2), any code-wrappers or bindings that are available (which allow for the algorithms written in one language to be accessed by another), and whether they are available as VAMP plug-ins.

| Name | Language | Initial Release | Runs in real-time? | Wrappers/Bindings | Vamp Plug-in |
|---|---|---|---|---|---|
| MARSYAS | C++ | 2000 | Yes | Python | X |
| Aubio | C | 2006 | Yes | Python | X |
| LibXtract | C | 2007 | Yes | Python | X |
| Essentia | C++ | 2013 | Yes | Python | - |
| jAudio | Java | 2005 | No | - | - |
| MIR Toolbox | MATLAB | 2007 | No | - | - |
| AMPACT | MATLAB | 2011 | No | - | - |
| Chroma Toolbox | MATLAB | 2011 | No | - | - |
| Timbre Toolbox | MATLAB | 2011 | No | - | - |
| LibROSA | Python | 2015 | Yes (using pyAudio) | - | - |
| MSAF | Python | 2015 | No | - | - |
| Madmom | Python | 2015 | Yes | - | - |

Table 3: Summary of the language the audio processing tools surveyed in this chapter are written and whether they are available as Vamp plug-ins.

A number of tools are written in C or C++, low-level, strongly typed languages that are used due to their performance speed and ability to work across platforms. C++ is also object-oriented, which facilitates the code reuse. C and C++ can, however, be hard for a novice user to work with, so developers often create wrappers for scripting languages to access C/C++ functions and compiled executables. Another advantage of C/C++ for audio developers is the ease with which developers can create VAMP plug-ins, since the VAMP Software Development Kit is written in C/C++. Thus, most of the C/C++ packages described below are also available as VAMP plug-ins.

Music Analysis, Retrieval and SYnthesis for Audio Signals (MARSYAS) is a software framework written in C++. It was originally released in 2000 (Tzanetakis and Cook 2000) and updated in 2008 (Tzanetakis 2008). MARSYAS can be run from the command line using its own scripting language (Marsyas Script), as well as with Python wrappers or as a VAMP plug-in. MARYSAS primarily extracts low-level features (including F0, chroma, MFCCs, spectral descriptors, and zero-crossing rate), although it also provides estimates of beat-locations and tempo. MARSYAS also has implementations of several statistical tools, including principal component analysis (PCA), self-organising maps (SOM), k-nearest neighbors (KNN), Gaussian mixture models (GMM), and support vector machines (SVM), which can be used to model the extracted features. Aubio and LibXtract, are written in C/C++ and are also implemented as VAMP plug-ins. Aubio is written in C (Brossier 2006) and performs low-level signal processing, including MFCCs, filters and phase vocoding (a technique for time-scaling audio without frequency shifts using phase information), as well as onset detection and beat and tempo tracking and pitch tracking, with several methods implemented for both onset detection and pitch tracking. LibXtract is written in C++ (Bullock 2007) and focuses on extracting low-level loudness, pitch, and timbral features.

Essentia is a more recent C++ library with python bindings that was developed by researchers at Universitat Pompeu Fabra (Bogdanov et al. 2013). Essentia includes low-level feature extraction and processing (including filters), statistical descriptors of the extracted low-level features, and a number of high-level features. These high-level features include melody, chord, key and tuning, onset, beat, and tempo estimation, note tracking, and segmentation. Essentia also includes some pre-trained classification models through integration with the Gaia 2 library for audio similarity and classification that researchers can use to assess similarity within and across audio files.

Java is an object-oriented cross-platform language like C++. While it lacks some of C++'s performance speed, it has easier memory management, making it more accessible for programmers. The most widely used audio processing tool in Java is jAudio (McKay, Fujinaga, and Depalle 2005, McEnnis et al. 2006), which provides a range of low-level loudness and timbral feature extractors as well as tools for combining low-level features into high-level features of the researcher's design. jAudio is part of the jMIR software suite (McKay 2010). jMIR includes tools for symbolic music analysis, lyric analysis, and mining musical data from the web, as well as ACE (autonomous classification engine). ACE (McKay et al. 2005) is an autonomous framework that searches for the optimal configuration of classifiers to suit a given task and set of features.

A number of audio-processing toolkits are written in MATLAB, due to the language's popularity for signal processing and machine learning from the early 2000's until a few years ago. MATLAB offers sophisticated visualisation tools as well as a range of powerful signal processing and statistical libraries. Machine learning tasks can be performed using either the built-in functionality or from third-party tools, like liblinear (Fan et al. 2008) for classification or the Probabilistic Modelling Toolkit (Murphy 2012) for both classification and temporal models. The first and one of the most widely used MATLAB toolkits is the MIR Toolbox (Lartillot and Toiviainen 2007). The MIR toolbox can extract a range of low- and high-level features related to loudness, pitch, timbre, and timing and also has some simple classification tools, specifically KNNs and GMMs. Recently, a beta of MiningSuite, an extension to the MIR Toolbox, was released, which integrates audio, symbolic, and video analysis. The Automatic Music Performance Analysis and Comparison Toolkit (AMPACT) aligns MIDI to audio for score-informed pitch, timing, dynamics and timbre information for both monophonic (Devaney et al. 2011) and polyphonic audio (Devaney and Mandel, 2017). The Chroma Toolbox (Müller and Ewert 2011) extracts a number of chroma-related features, including energy-normalised chroma and estimates of musical notes. The Chroma Toolbox is useful for generating low-level features that can subsequently be used for chord estimation, structural analysis, or music alignment. The Timbre Toolbox (Peeters et al. 2011) extracts a range of global and time-varying timbre-related features based on decades of research into timbre perception.

Python has become a significant programming language for audio processing over the past few years, which is reflected in the number of audio processing Python packages released recently. Audio processing packages build on general numerical processing packages, particularly scipy (Jones, Oliphant, and Peterson 2001), numpy (Oliphant 2006), and matplotlib (Hunter 2007) and provide specific music audio processing functions. LibROSA (McFee et al. 2015) provides Python functions for analysing and visualising musical audio. The analysis functions perform both lower-level tasks, like spectrogram decomposition, filtering, and feature extraction, and higher-level tasks like onset detection, beat and tempo estimation, and segmentation. It also includes some tools for modelling time-series data. The Music Structure Analysis Framework (MSAF) (Nieto and Bello 2016) provides a set of segmentation algorithms that either automatically detect section boundaries or automatically cluster labels. It also has lower-level feature extraction functions as well algorithm evaluation functions. Madmom is a Python audio signal processing library (Böck et al. 2016) that extracts a range of higher-level features including onset detection, beat detection, downbeat detection, note segmentation, chord estimation, and key estimation. Madmom also includes implementations of several temporal machine learning algorithms: conditional random fields, Gaussian mixture models, hidden Markov models, and shallow neural networks.

## 4 Conclusions

While audio processing tools have made great advances over the last fifteen years, they are still not sufficiently accurate to transcribe or describe music in a fully automatic manner. Thus, they typically either require manual intervention or the design of subsequent analyses that allow for some degree of error in the information extracted from the audio. As such, a large number of musicologists and theorists still make use of manual annotation for these types of studies, facilitated by software with graphical user interfaces, such as Audacity or Sonic Visualiser. One part of the challenge of improving automatic processing is accumulating enough labelled data, which could be facilitated by the sharing of manual annotations. Another part of the challenge is dealing with differences between human annotators, which is particularly problematic for higher-level features. For lower-level features, there is usually a perceptual threshold at which most people agree, such as the 20 cent threshold for the just noticeable difference in the pitch of musical notes (Vurma and Ross 2006). For higher-level features, such as chord estimation (Humphrey and Bello 2015), similarity (Flexer 2015), or structural segmentation (Smith et al. 2011) there are often significant amounts of disagreement in human-generated annotations themselves, as the tasks themselves are poorly defined. This could potentially also be addressed with more annotated data. (Much) larger amounts of annotated data would also make deep learning approaches more viable, which, based on the successes seen in the speech community, could improve overall performance on tasks such as transcription. One challenge with deep learning approaches, as with all machine learning approaches, is that they typically cannot provide an accurate confidence level in their predictions, meaning it is difficult for researchers to know when to distrust their answers. In the near term, machines may not be able to replace human annotation entirely, but the ability to study much larger corpora of audio recordings, even with some annotation error, is a still a worthwhile trade-off for many research questions.